\documentclass{ifacconf}

\usepackage{graphicx}      
\usepackage{natbib}        

\usepackage{xcolor}
\usepackage{amsmath,amssymb,amsfonts}
\usepackage{algorithmic}
\usepackage{graphicx}
\usepackage{textcomp}

\usepackage{mathtools}
\usepackage[acronym, style=alttree, toc=true, shortcuts, xindy, nomain, nonumberlist]{glossaries}
\usepackage{hyphenat}
\usepackage{bm}
\usepackage{siunitx}  

\usepackage{IEEEtrantools}

\newcommand{\ma}[1]{\bm{#1}}

\newcommand{\norm}[1]{\left\lVert#1\right\rVert} 
\newcommand{\abs}[1]{\left|#1\right|} 
\newcommand{\set}[1]{\mathcal{#1}} 
\newcommand{\argmin}[1]{\underset{#1}{\hspace{.1cm}\mathrm{arg}\hspace{.05cm}\mathrm{min}} \hspace{.1cm}} 

\newcommand{\interior}[1]{\text{int}\left(#1\right)} 
 
\newcommand{\ol}[1]{\overline{#1}}
\newcommand{\mat}[1]{\ensuremath{\begin{bmatrix} #1 \end{bmatrix}}}				

\DeclarePairedDelimiterXPP\onenorm[1]{}\lVert\rVert{_1}{\ifblank{#1}{\:\cdot\:}{#1}} 
\DeclarePairedDelimiterXPP\twonorm[1]{}\lVert\rVert{_2}{\ifblank{#1}{\:\cdot\:}{#1}} 

\newtheorem{assumption}{Assumption}

\newglossary{symbols}{sym}{sbl}{List of Symbols}
\newglossary{notation}{not}{nt}{Notation}

\glsaddkey{dimension}{\glsentrytext{\glslabel}}{\glsentryd}{\GLsentryd}{\glsd}{\Glsd}{\GLSd}
\glsaddkey{body}{\glsentrytext{\glslabel}}{\glsentrybody}{\GLsentrybody}{\glsbody}{\Glsbody}{\GLSbody}

\newacronym{MPC}{MPC}{Model Predictive Control}
\newacronym{RMPC}{RMPC}{Robust Model Predictive Control}
\newacronym{SMPC}{SMPC}{Stochastic Model Predictive Control}
\newacronym{SCMPC}{SCMPC}{Scenario Model Predictive Control}
\newacronym{OCP}{OCP}{optimal control problem}

\newglossaryentry{xref}{type=symbols,
	sort={general},
	name={\ensuremath{\bm{x}_\text{ref}}},
	description={x reference}
}

\begin{document}
\begin{frontmatter}

\title{Combined Robust and Stochastic Model Predictive Control for Models of Different Granularity\thanksref{footnoteinfo}} 

\thanks[footnoteinfo]{The authors gratefully acknowledge the financial and scientific support by the BMW Group within the CAR@TUM project.}

\author[First]{Tim Br\"udigam} 
\author[First]{Johannes Teutsch} 
\author[First]{Dirk Wollherr}
\author[First]{Marion Leibold}

\address[First]{Chair of Automatic Control Engineering, Technical University of Munich, Arcisstr. 21, 80333 Munich, Germany (email: \{tim.bruedigam;~johannes.teutsch;~dw;~marion.leibold\}@tum.de)}

\begin{abstract}                
Long prediction horizons in Model Predictive Control (MPC) often prove to be efficient, however, this comes with increased computational cost. Recently, a Robust Model Predictive Control (RMPC) method has been proposed which exploits models of different granularity. The prediction over the control horizon is split into short-term predictions with a detailed model using MPC and long-term predictions with a coarse model using RMPC. In many applications robustness is required for the short-term future, but in the long-term future, subject to major uncertainty and potential modeling difficulties, robust planning can lead to highly conservative solutions. We therefore propose combining RMPC on a detailed model for short-term predictions and Stochastic MPC (SMPC), with chance constraints, on a simplified model for long-term predictions. This yields decreased computational effort due to a simple model for long-term predictions, and less conservative solutions, as robustness is only required for short-term predictions. The effectiveness of the method is shown in a mobile robot collision avoidance simulation.
\end{abstract}


\end{frontmatter}

\section{Introduction}
\label{sec:introduction}

\vspace{-15cm}
\mbox{\small \textcopyright~2020~the~authors.~This~work~has~been~accepted~to~IFAC~for~publication~under~a~Creative~Commons~Licence~CC-BY-NC-ND.}
\vspace{14.2cm}

\vspace{-14.7cm}
\mbox{\small The~published~version~may~be~found~at~https://doi.org/10.1016/j.ifacol.2020.12.515.}
\vspace{13.7cm}

\gls{MPC} iteratively solves an optimal control problem on a finite horizon, given the prediction of the system behavior based on a system model. While precise models decrease errors in the prediction, the computational effort increases. In stochastic environments even precise models contain random variables to account for the present system uncertainty. Long prediction horizons allow to consider system behavior further in the future, but they also increase conservatism as the uncertainty increases. All three issues - model precision, environment uncertainty, and conservatism – are fundamental for the efficient application of \gls{MPC}.\\
\gls{MPC} with different prediction models for varying process parts was studied in \cite{Lu2015} and \cite{FarinaZhangScattolini2018}. While these works use separate optimization problems for individual processes, in \cite{BaethgeLuciaFindeisen2016} a method is suggested which uses models of different granularity within one \gls{MPC} optimal control problem. The prediction horizon is split into a short-term prediction with a detailed model and a long-term prediction with a coarse model. This approach allows for precise control in the immediate future, while still considering a longer horizon.\\ 
Environment uncertainty within \gls{MPC} is addressed by \gls{RMPC} \citep{Mayne2014}. Considering the worst-case uncertainty realization, \gls{RMPC} provides a control law which ensures robustness. However, robustly satisfying constraints with \gls{RMPC} can result in overly conservative solutions, leading to \gls{SMPC} \citep{Mesbah2016, FarinaGiulioniScattolini2016}. In \gls{SMPC} chance constraints are applied, allowing a small, predefined level of constraint violation, which reduces conservatism while risk is increased. Among the varying approaches to \gls{SMPC} are tube-based \gls{SMPC} \citep{KouvaritakisEtalCheng2010,CannonEtalCheng2011}, \gls{SCMPC} in \cite{SchildbachEtalMorari2014}, and a combination of \gls{SMPC} and \gls{SCMPC} in \cite{BruedigamEtalWollherr2018b}.\\
In this paper we propose combining models of different granularity with \gls{RMPC} and \gls{SMPC}. For brevity we will refer to the proposed method as granularity R+SMPC. The prediction horizon and the optimal control problem are split into two parts. \gls{RMPC} uses a detailed model for short-term predictions, while \gls{SMPC} is applied for the long term, making predictions with a coarse model. This coarse model can be an approximated model compared to the detailed model applied for the short-term prediction.\\
\gls{RMPC} making predictions with a detailed model for the short term ensures that constraints are satisfied in the presence of uncertainty. For the short term, the overall error of the prediction is reduced by a detailed model, due to a small modeling error and manageable system uncertainty. For long-term predictions the prediction error increases. This prediction error is increasingly influenced by the propagated system uncertainty, decreasing the benefit of applying a detailed prediction model. Therefore, a less detailed, coarse model is used for long-term predictions to reduce computational complexity. Applying \gls{RMPC} for the long term would result in a conservative solution of the optimal control problem. Therefore, \gls{SMPC} with chance constraints is used for the long term to reduce conservatism, as precise and robust control actions are often not sensible for long-term planning.\\
The proposed method allows to plan robustly for the immediate future, while still considering a longer horizon without overly restrictive solutions due to increased system uncertainty. This approach can be beneficial in safety-critical applications such as autonomous driving \citep{BruedigamEtalWollherr2018b, CarvalhoEtalBorrelli2014, PekAlthoff2018}, where collision avoidance must be ensured in the immediate future, while considering a longer horizon allows to plan efficiently. For example, decelerating before a turn can be performed more smoothly the earlier the prediction includes the turn, resulting in more comfort for passengers. However, planning robustly for long horizons is impractical as the prediction of traffic participant behavior is imprecise, especially for pedestrians in urban scenarios \citep{KoschiEtalAlthoff2018}, resulting in overly conservative trajectory planning. Applying the proposed method addresses both safe planning as well as considering long horizons with reduced computational effort.

This paper is structured as follows. Section \ref{sec:problem} introduces the problem setup, the proposed method is described in Section \ref{sec:method}. Section \ref{sec:results} illustrates the application of the proposed method in a simulation example, while a conclusion is given in Section~\ref{sec:conclusion}.

\section{Problem Setup}
\label{sec:problem}

We consider two models of different granularity for a linear, discrete time system with additive disturbance
\begin{IEEEeqnarray}{rl}
\IEEEyesnumber \label{eq:systemmodels}
\bm{x}_{k+1} &= \bm{A} \bm{x}_k+\bm{B} \bm{u}_k + \bm{d}_k, \IEEEyessubnumber \label{eq:detailed}\\
\bm{\xi}_{k+1} &= \bm{A}_{\text{c}} \bm{\xi}_k+\bm{B}_{\text{c}} \bm{v}_k + \bm{G}_{\text{c}} \bm{w}_k, \IEEEyessubnumber \label{eq:coarse}
\end{IEEEeqnarray}
where $\bm{x}_k \in \mathbb{R}^{n_x}$ and $\bm{\xi}_k \in \mathbb{R}^{n_{\xi}}$ denote the states, and $\bm{u}_k \in \mathbb{R}^{n_u}$ and $\bm{v}_k \in \mathbb{R}^{n_v}$ the inputs at the time step $k$, and $\bm{A} \in \mathbb{R}^{n_{x} \times n_{x}}$, $\bm{B} \in \mathbb{R}^{n_{x} \times n_{u}}$, $\bm{A}_\text{c} \in \mathbb{R}^{n_{\xi} \times n_{\xi}}$, $\bm{B}_\text{c} \in \mathbb{R}^{n_{\xi} \times n_{v}}$, $\bm{G}_{\text{c}} \in \mathbb{R}^{n_{\xi} \times n_{\xi}}$. The bounded disturbance $\bm{d}_k \in \mathbb{D} \subset \mathbb{R}^{n_x}$ denotes additive uncertainty within the system model, where $\mathbb{D}$ is a compact convex set and includes the origin. While \eqref{eq:detailed} is a more detailed model, \eqref{eq:coarse} is a coarse representation of system \eqref{eq:detailed} with an additional normally distributed, zero mean random variable $\bm{w}_k \sim \mathcal{N}\left(\bm{0}, \bm{\Sigma}^{w}\right)$ with covariance matrix $\bm{\Sigma}^{w}$. Thus, $\bm{w}_k$ is unbounded and can be chosen as an over-approximation of the bounded disturbance $\bm{d}_k$ from the detailed model.\\
The states and inputs of both models are constrained by
\begin{equation}
\bm{x}_k \in \mathbb{X},\hspace{3pt} \bm{u}_k \in \mathbb{U},\hspace{3pt} \bm{\xi}_k \in \Xi,\hspace{3pt} \bm{v}_k \in \mathbb{V} \hspace{10pt} \forall k \in \mathbb{N}. \label{eq:systemconstraints}
\end{equation}
To connect these models within the prediction and to ensure consistency between the models, the following projection is used, similar to \cite{BaethgeLuciaFindeisen2016}.
\begin{assumption}[Projection]\label{ass:proj}
(a) There exists a surjective projection function $\text{Proj}: \mathbb{R}^{n_{x}} \times \mathbb{R}^{n_{u}} \rightarrow \mathbb{R}^{n_{\xi}} \times \mathbb{R}^{n_{v}}$, which maps the states $\bm{x}_k$ and inputs $\bm{u}_k$ of the detailed model~\eqref{eq:detailed} to the states $\bm{\xi}_k$ and inputs $\bm{v}_k$ of the coarse model~\eqref{eq:coarse}, i.e., $(\bm{\xi}_k, \bm{v}_k) = \text{Proj}\left(\bm{x}_k,\bm{u}_k\right)$.\\
(b) The constraint sets $\Xi$ and $\mathbb{V}$ of the coarse model can be computed by projecting the constraint sets $\mathbb{X}$ and $\mathbb{U}$ of the detailed model, i.e., $(\Xi, \mathbb{V}) = \text{Proj}\left(\mathbb{X},\mathbb{U}\right)$.
\end{assumption}
The two models and the corresponding projection function are now used to design an MPC method with models of different granularity, applying robust constraints for the immediate future and chance constraints for long-term predictions.

\section{\gls{RMPC} and \gls{SMPC} with Models of Different Granularity}
\label{sec:method}

\begin{figure}
	\includegraphics[width=\columnwidth,trim=0cm 0.8cm 0cm 0.5cm]{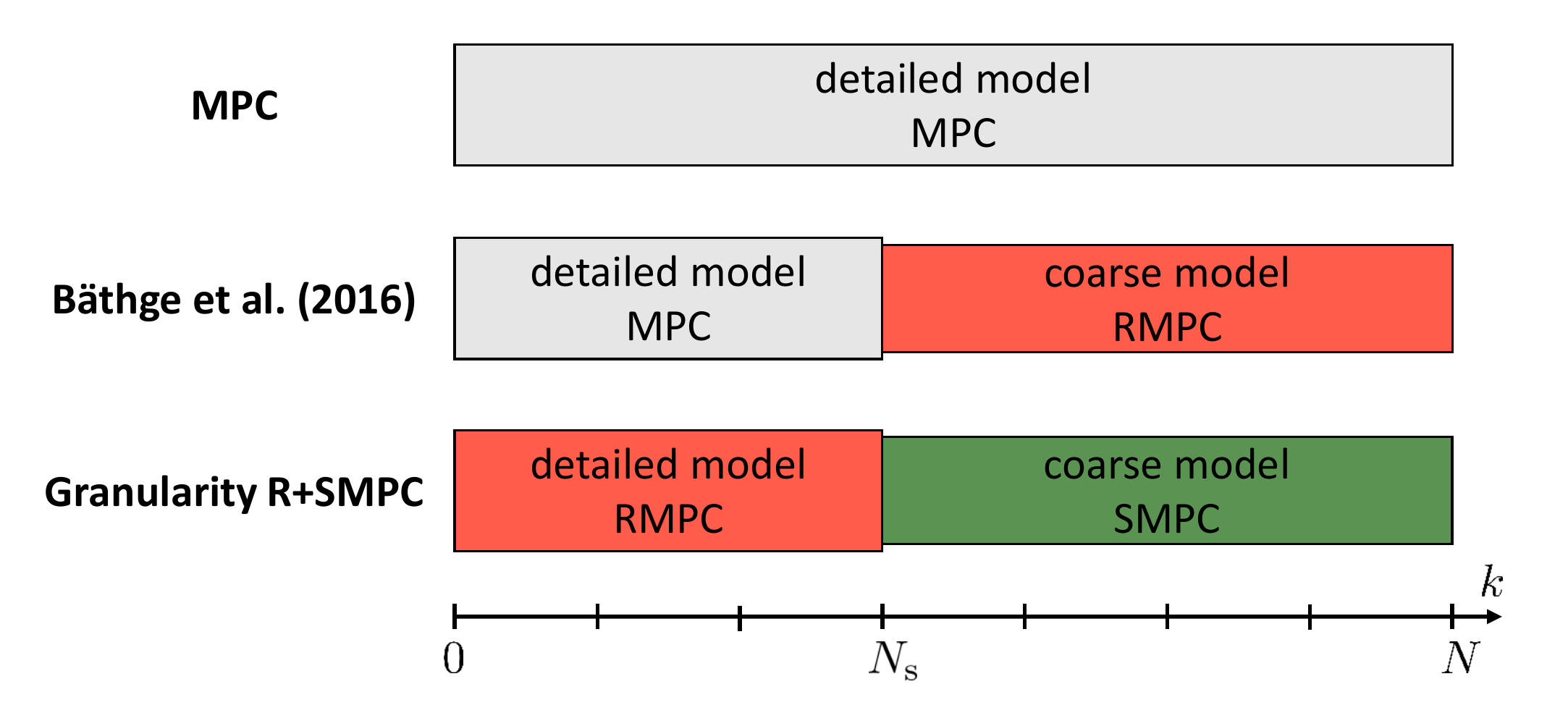}
	\caption[Method]{Comparison of granularity R+SMPC to MPC and \cite{BaethgeLuciaFindeisen2016}.}
	\label{fig:method}
\end{figure}

In the following, an MPC \gls{OCP} is derived with RMPC and a detailed model for short-term predictions, as well as SMPC with chance constraints and a coarse model for long-term predictions. The proposed method is referred to as granularity R+SMPC. We will first present the general structure of the proposed granularity R+SMPC \gls{OCP}. Then, details are provided on the robust constraints and chance constraints of RMPC and SMPC, respectively. Eventually, the resulting overall granularity R+SMPC \gls{OCP} is shown. The proposed method is illustrated in Figure \ref{fig:method}.

We first formulate an \gls{MPC} routine with two prediction stages and a total of $N$ prediction steps. The detailed model \eqref{eq:detailed} is used for the short-term prediction, with $N_\text{s}$ prediction steps, to guarantee robust constraint satisfaction in the immediate future. Additionally, the coarse model \eqref{eq:coarse} is used for the long-term prediction, with $N_\text{l}~=~N~-~N_\text{s}$ prediction steps, to reduce the computational cost while considering the uncertainties in a probabilistic manner by turning the state constraints into chance constraints. The \gls{OCP} is given by 
\begin{IEEEeqnarray}{rl}
\IEEEyesnumber \label{eq:generalocp}
	\bm{U}^* &= \argmin{\bm{U}}  \sum\limits_{k=0}^{N_{\text{s}}-1} \text{E}\left[l \left(\bm{x}_k,\bm{u}_k\right)\right] \notag \\ & \hspace{14mm}+ \sum\limits_{k=N_{\text{s}}}^{N-1} \text{E}\left[l_{\text{c}} \left(\bm{\xi}_k,\bm{v}_k\right)\right] + \text{E}\left[V_{\text{f,c}}\left(\bm{\xi}_{N}\right)\right] \IEEEyessubnumber \\
	\text{s.t. } & \bm{x}_{k+1} = \bm{A} \bm{x}_k + \bm{B} \bm{u}_k +  \bm{d}_k, \IEEEyessubnumber \\
	& \bm{x}_k \in \mathbb{X} \hspace{43pt}~~\forall k \in \{0,\dots,N_{\text{s}}\} , \IEEEyessubnumber\\
	& \bm{u}_k \in \mathbb{U} \hspace{43pt}~~\forall k \in \{0,\dots,N_{\text{s}}-1\} ,\IEEEyessubnumber\\
	& \left(\bm{\xi}_{N_{\text{s}}}, \bm{v}_{N_{\text{s}}}\right) = \text{Proj}\left(\bm{x}_{N_{\text{s}}}, \bm{u}_{N_{\text{s}}}\right), \IEEEyessubnumber \label{eq:generalocp_proj}\\
	& \bm{\xi}_{k+1} = \bm{A}_{\text{c}} \bm{\xi}_k + \bm{B}_{\text{c}} \bm{v}_k + \ma{G}_\text{c} \bm{w}_k, \label{coarse} \IEEEyessubnumber\\
	& \text{Pr}\left(\bm{\xi}_k \in \Xi\right) \ge p \hspace{8pt}~~\forall k \in \{N_{\text{s}},\dots,N\}, \IEEEyessubnumber \label{probcon}\\
	& \bm{v}_k \in \mathbb{V} \hspace{52pt} \forall k \in \{N_{\text{s}},\dots,N-1\}, \IEEEyessubnumber
\end{IEEEeqnarray}

with the input sequence $\bm{U} = \left(\bm{u}_0,\dots,\bm{u}_{N_{\text{s}}},\bm{v}_{N_{\text{s}}},\dots,\bm{v}_{N}\right)^\top$, the running cost functions $l: \mathbb{R}^{n_{x}} \times \mathbb{R}^{n_{u}} \rightarrow \mathbb{R}_+ $ and $l_{\text{c}}:~\mathbb{R}^{n_{\xi}}~\times~\mathbb{R}^{n_{v}}~\rightarrow~\mathbb{R}_+ $, as well as the terminal cost function ${ V_{\text{f,c}}:\mathbb{R}^{n_{\xi}} \rightarrow \mathbb{R}_+ }$. The risk parameter $p$ in \eqref{probcon} specifies the desired probability of state constraint satisfaction in the long-term prediction, i.e., the probability of violating the constraint in each prediction step is $1-p$. Note that $\bm{u}_{N_\text{s}}$ is necessary only to compute \eqref{eq:generalocp_proj}. \\
In the following, we modify the \gls{OCP}~\eqref{eq:generalocp} such that we can guarantee robust constraint satisfaction for the short-term prediction considering any possible disturbance sequence $\bm{D}_k = \left(\bm{d}_0,\dots,\bm{d}_k\right)^{\top}$. It is also necessary to reformulate the probabilistic constraint \eqref{probcon} into a deterministic expression, such that it is tractable for the solver.

\subsection{RMPC with a Detailed Model}
For the short-term prediction with robust constraints a disturbance-free reference system is defined, also called nominal system, which has tighter constraints \citep{RawlingsMayneDiehl2017}. First, we decompose the input into a stabilizing state feedback and a new decision variable $\bm{\nu}_k$ for the controller, i.e.,
\begin{equation}
    \bm{u}_k = \bm{K} \bm{x}_k + \bm{\nu}_k.
\end{equation}
with feedback gain $\bm{K}$. The actual system model and the nominal system model are given by
\begin{IEEEeqnarray}{rl}
\IEEEyesnumber \label{eq:moddetsys}
\bm{x}_{k+1} &= \bm{\Phi} \bm{x}_k + \bm{B} \bm{\nu}_k + \bm{d}_k, \IEEEyessubnumber \label{eq:detsys}\\
\ol{\bm{x}}_{k+1} &= \bm{\Phi} \ol{\bm{x}}_k + \bm{B} \bm{\nu}_k,\IEEEyessubnumber \label{eq:nominalsys}
\end{IEEEeqnarray}
with the stabilized system matrix $\bm{\Phi} = \bm{A} + \bm{B} \bm{K}$, and the nominal states $\ol{\bm{x}}_k$. By subtracting \eqref{eq:nominalsys} from \eqref{eq:detsys}, we derive an equation for the deviation $\bm{\varepsilon}_k := \bm{x}_k - \ol{\bm{x}}_k$ of the actual and nominal state, i.e.,
\begin{equation}
    \bm{\varepsilon}_{k+1} = \bm{\Phi} \bm{\varepsilon}_k + \bm{d}_k. \label{eq:erroreq}
\end{equation}
\begin{assumption} \label{ass_e0}
There is no deviation between the actual state and the nominal state at time instant $k=0$, i.e.,
\begin{equation}
    \bm{\varepsilon}_0 = 0 \hspace{3pt} \Leftrightarrow \hspace{3pt} \bm{x}_0 = \ol{\bm{x}}_0.
\end{equation}
\end{assumption}
With Assumption \ref{ass_e0} and the Minkowski set addition, we determine the set containing $\bm{\varepsilon}_k$
\begin{equation}
    \mathbb{S}_{k} := \bigoplus\limits_{i=0}^{k-1} \bm{\Phi}^i \mathbb{D} = \mathbb{D} \oplus \bm{\Phi} \mathbb{D} \oplus \dots \oplus \bm{\Phi}^{k-1} \mathbb{D}.
\end{equation}
It is now possible to compute the minimal disturbance invariant set $\mathbb{Z} := \mathbb{S}_{\infty}$, which is used to define an outer-bounding tube around the states of the nominal system $\ol{\bm{x}}_k$ in which the states of the actual system $\bm{x}_k$ lie for any possible disturbance sequence $\bm{D}_k$, i.e.,
\begin{equation}
    \bm{x}_k \in \{\ol{\bm{x}}_k\} \oplus \mathbb{Z}. \label{eq:tube}
\end{equation}
Tighter constraint sets for states and inputs of the nominal system are now computed with \eqref{eq:tube}, resulting in
\begin{IEEEeqnarray}{rl}
\IEEEyesnumber \label{eq:tightcons}
\ol{\mathbb{X}} &= \mathbb{X} \ominus \mathbb{Z}, \IEEEyessubnumber\\
\ol{\mathbb{U}} &= \mathbb{U} \ominus \ma{K} \mathbb{Z}, \IEEEyessubnumber
\end{IEEEeqnarray}
under the condition that the set $\mathbb{D}$ is small enough to ensure that $\mathbb{Z} \subset \interior{\mathbb{X}}$ and $\ma{K} \mathbb{Z} \subset \interior{\mathbb{U}}$ hold.

\subsection{Improved RMPC Optimal Control Problem}\label{sec:impRMPC}
Although using the nominal system \eqref{eq:nominalsys} and the tightened constraint sets \eqref{eq:tightcons} for the short-term prediction would lead to robust constraint satisfaction, it is possible to improve this method by making use of an additional degree of freedom to the controller, namely the initial state of the nominal system $\ol{\bm{x}}_0$ \citep{MayneSeronRakovic2005}. There is no guarantee that setting the initial state of the nominal system $\ol{\bm{x}}_0$ equal to the actual initial state $\bm{x}_0$ enhances convergence to the reference of the nominal state trajectory. In order to determine an improved center of the tube, the controller considers the initial state of the nominal system $\ol{\bm{x}}_0$ as an additional decision variable to the inputs $\bm{\nu}_k$. It is necessary that the actual initial (current) state $\bm{x}_0$ remains in the tube with the initial nominal state $\ol{\bm{x}}_0$ as its center, i.e.,
\begin{equation}
	\bm{x}_0 \in \{\ol{\bm x}_0\} \oplus \mathbb{Z} \hspace{3pt} \Leftrightarrow  \hspace{3pt} \bm{x}_0 - \ol{\bm x}_0 \in \mathbb{Z},
\end{equation} which is treated as a constraint for this decision variable. This method yields faster convergence and additionally has pleasing theoretical properties considering stability \citep{MayneSeronRakovic2005, RawlingsMayneDiehl2017}.

\subsection{SMPC with a Coarse Model}\label{sec:smpc}
The probabilistic constraint \eqref{probcon} needs to be reformulated into a deterministic expression, as shown in \cite{CarvalhoEtalBorrelli2014}, to implement the Stochastic MPC scheme on the detailed model for the long-term prediction. Therefore, we first determine the uncertainty propagation within the coarse model by decomposing the states $\bm{\xi}_k$ into a deterministic and a probabilistic component, and the inputs $\bm{v}_k$ into a stabilizing state feedback and a new decision variable for the controller, i.e.,
\begin{IEEEeqnarray}{rl}
\IEEEyesnumber \label{eq:decomp}
\bm{\xi}_k &= \bm{z}_k + \bm{e}_k, \IEEEyessubnumber\\
\bm{v}_k &= \bm{K}_{\text{c}} \bm{\xi}_k + \bm{c}_k. \IEEEyessubnumber
\end{IEEEeqnarray}
Substituting for $\bm{\xi}_k$ and $\bm{v}_k$ in the system equation yields
\begin{IEEEeqnarray}{rl}
\IEEEyesnumber \label{eq:decomp}
\bm{z}_{k+1} &= \bm{\Phi}_{\text{c}} \bm{z}_k + \bm{B}_{\text{c}} \bm{c}_k, \IEEEyessubnumber\\
\bm{e}_{k+1} &= \bm{\Phi}_{\text{c}} \bm{e}_k + \bm{G}_{\text{c}} \bm{w}_k, \IEEEyessubnumber \label{eq:decomp_prob} 
\end{IEEEeqnarray}
with the stabilized system matrix $\bm{\Phi}_{\text{c}} = \bm{A}_{\text{c}} + \bm{B}_{\text{c}} \bm{K}_{\text{c}}$.
We now determine the distribution of the probabilistic component $\bm{e}_k$ for $k>0$, where the distributions of $\bm{e}_0$ and $\bm{w}_k$ are known. Due to the normally distributed, zero mean random disturbance $\bm{w}_k$, $\bm{e}_k$ is also normally distributed and zero mean, with the covariance matrix~$\bm{\Sigma}_k^e$. The uncertainty propagation is computed using \eqref{eq:decomp_prob}, which yields 
\begin{equation}
    \bm{\Sigma}_{k+1}^e = \bm{\Phi}_{\text{c}} \bm{\Sigma}_k^e \bm{\Phi}_{\text{c}}^\top + \bm{G}_{\text{c}} \bm{\Sigma}^w \bm{G}_{\text{c}}^\top. \label{eq:propagation}
\end{equation}
Due to $\bm{w}_k$ being an over-approximation of the disturbance in the detailed model, it follows from Assumption \ref{ass_e0} that $\bm{e}_0 = 0$, and thus $\bm{\Sigma}_0^e = \bm{0}$. Note that the coarse model is used starting from time instant $N_{\text{s}}$, so $\bm{\Sigma}_{N_{\text{s}}}^e$ has to be pre-computed via \eqref{eq:propagation}. \\
In order to reformulate the chance constraint, we first assume that it is possible to describe the state constraint $\bm{\xi}_k \in \Xi$ by a function $g_k$, i.e.,
\begin{equation}
    g_k(\bm{\xi}_k) \ge 0 \Leftrightarrow \bm{\xi}_k \in \Xi,
\end{equation}
so that the state constraint is satisfied if $g_k \ge 0$ and violated if $g_k < 0$. In general, it is also possible to have multiple chance constraints, i.e., multiple inequality constraints which describe the constraint set $\Xi$. We show the procedure for a single chance constraint, which can then be used for every additional chance constraint.
Generally, $g_k$ depends nonlinearly on the state $\bm{\xi}_k$. After linearizing $g_k$ around the predicted (deterministic) states $\bm{z}_k$, the state constraint becomes 
\begin{equation}
    g_k(\bm{z}_k) + \nabla g_k (\bm{z}_k)^\top \bm{e}_k \ge 0,
\end{equation}
where $\nabla g_k = \frac{\partial g_k}{\partial \bm{\xi}_k}$ is the gradient of $g_k$. Thus, the chance constraint can be substituted by 
\begin{equation}
    \text{Pr}\left(-\nabla g_k (\bm{z}_k)^\top \bm{e}_k \le g_k(\bm{z}_k)\right) \ge p.
\end{equation}
We now split this constraint into a deterministic inequality and a probabilistic equation
\begin{IEEEeqnarray}{rl} 
\IEEEyesnumber \label{eq:probeq}
&g_k(\bm{z}_k) \ge \gamma_k, \IEEEyessubnumber\\
&\text{Pr}\left(-\nabla g_k (\bm{z}_k)^\top \bm{e}_k \le \gamma_k \right) = p.\IEEEyessubnumber
\end{IEEEeqnarray}
As $\bm{e}_k \sim \mathcal{N}\left(\bm{0}, \bm{\Sigma}_k^e\right)$, it follows that
\begin{equation}
-{\nabla g_k}^T\bm{e}_k \sim \mathcal{N} \left(\bm{0}, {{\nabla g_k}^\top\bm{\Sigma}_k^e{\nabla g_k}}\right),
\end{equation}
and \eqref{eq:probeq} is solved for the parameter $\gamma_k$ using the quantile function for univariate normal distributions. This yields the deterministic expression of the chance constraint, i.e.,
\begin{IEEEeqnarray}{rl} \label{eq:probeq_analytic}
\IEEEyesnumber
&g_k(\bm{z}_k) \ge \gamma_k, \IEEEyessubnumber\\
&\gamma_k = \sqrt{2{\nabla g_k(\bm{z}_k)}^\top\bm{\Sigma}_k^e{\nabla g_k(\bm{z}_k)}}\text{ erf}^{-1}\left(2p-1\right).\IEEEyessubnumber
\end{IEEEeqnarray}
With this approach a deterministic expression of the probabilistic chance constraint is provided.

\subsection{Granularity R+SMPC Optimal Control Problem}

Using the results from this section for the \gls{OCP} \eqref{eq:generalocp}, the overall granularity R+SMPC \gls{OCP} is given by
\begin{IEEEeqnarray}{rl}
\IEEEyesnumber \label{eq:modifiedocp}
	\left(\ol{\bm{x}}_{0}^*,\bm{V}^*\right) =& \argmin{\ol{\bm{x}}_0,\bm{V}}  \sum\limits_{k=0}^{N_{\text{s}}-1} l \left(\ol{\bm{x}}_k,\ma{K}\ol{\bm{x}}_k + \set{\nu}_k\right) \notag \\ & \hspace{2mm}+ \sum\limits_{k=N_{\text{s}}}^{N-1} l_{\text{c}}\left(\bm{z}_k,\bm{K}_{\text{c}} \bm{z}_k + \bm{c}_k\right) +  V_{\text{f,c}}\left(\bm{z}_{N}\right) \IEEEyessubnumber \IEEEeqnarraynumspace \\
	\text{s.t. } & \bm{x}_0 - \ol{\bm x}_0 \in \mathbb{Z}, \IEEEyessubnumber \\
	& \ol{\bm{x}}_{k+1} = \bm{\Phi}  \ol{\bm{x}}_k + \bm{B}  \bm{\nu}_k,  \IEEEyessubnumber \\
	& \ol{\bm{x}}_k \in \ol{\mathbb{X}} \hspace{42pt}~~\forall k \in \{0,\dots,N_{\text{s}}\} , \IEEEyessubnumber\\
	& \ma{K}\ol{\bm{x}}_k + \bm{\nu}_k \in \ol{\mathbb{U}} \hspace{12pt}~\forall k \in \{0,\dots,N_{\text{s}}-1\} ,\IEEEyessubnumber\\
	& \left(\bm{z}_{N_{\text{s}}}, \bm{v}_{N_{\text{s}}}\right) = \text{Proj}\left(\ol{\bm{x}}_{N_{\text{s}}}, \bm{K}\ol{\bm{x}}_{N_{\text{s}}} + \bm{\nu}_{N_{\text{s}}}\right), \IEEEyessubnumber\\
	& \bm{c}_{N_{\text{s}}} = \bm{v}_{N_{\text{s}}} - \bm{K}_{\text{c}} \bm{z}_{N_{\text{s}}}, \IEEEyessubnumber\\
	& \bm{z}_{k+1} = \bm{\Phi}_{\text{c}}  \bm{z}_k + \bm{B}_{\text{c}}  \bm{c}_k,  \IEEEyessubnumber\\
	& g_k \ge \gamma_k \hspace{40pt}~~\forall k \in \{N_{\text{s}},\dots,N\}, \IEEEyessubnumber \\
	& \gamma_k = \sqrt{2{\nabla g_k}^\top\ma{\Sigma}_k^e{\nabla g_k}}\text{ erf}^{-1}\left(2p-1\right), \IEEEyessubnumber \\
	& \bm{K}_{\text{c}}\bm{z}_k + \bm{c}_k \in \mathbb{V} \hspace{15pt} \forall k \in \{N_{\text{s}},\dots,N-1\}, \IEEEyessubnumber \IEEEeqnarraynumspace
\end{IEEEeqnarray}
with the input sequence $\bm{V} = \left(\bm{\nu}_0,\dots,\bm{\nu}_{N_{\text{s}}},\bm{c}_{N_{\text{s}}},\dots,\bm{c}_{N}\right)^\top$.
Similar to \cite{MayneSeronRakovic2005}, the feedback control law $\bm{\kappa}^*(\cdot)$, which results from the solution of the above stated \gls{OCP}~\eqref{eq:modifiedocp}, follows
\begin{equation}
	\bm{\kappa}^*(\bm{x}_0) := \ol{\bm{u}}_0^* + \bm{K}\left(\bm{x}_0-\ol{\bm{x}}_0^*\right).
\end{equation}
This can be simplified to
\begin{equation} \label{control_rob}
	\bm{\kappa}^*(\bm{x}_0) = \bm{K}\ol{\bm{x}}_0^* + \bm{\nu}_0^* + \bm{K}\left(\bm{x}_0-\ol{\bm{x}}_0^*\right) = \bm{K}\bm{x}_0 + \bm{\nu}_0^*,
\end{equation}
where $\bm{\nu}_0^*$ is the first element of the optimal control sequence $\bm{V}^*$ and $\bm{x}_0$ is the actually sampled system state.

\subsection{Discussion}
\label{sec:discussion}

The proposed granularity R+SMPC method allows to robustly plan for a short-term horizon and consider long-term targets. Applying the chance constraint instead of a robust constraint for the long-term prediction reduces conservatism, as robustly accounting for uncertainties over a long horizon is often highly restrictive. Additionally, using the coarse model decreases the computational effort, which is the disadvantage of longer prediction horizons. It is to note that it is possible to combine more than two models of different granularity.\\
The projection mentioned in Assumption \ref{ass:proj} can be challenging to obtain, especially for more complex systems. Assuming the coarse model is an approximation of the detailed system, an approximation can also be used to define an appropriate system uncertainty for the coarse system, given the uncertainty in the detailed model. How to determine suitable coarse models and the corresponding uncertainties is a topic for further study.  \\
In \cite{BaethgeLuciaFindeisen2016} recursive feasibility of the MPC method with models of different granularity is proved with robust constraints for long-term predictions. Recursive feasibility guarantees that the MPC \gls{OCP} remains solvable in the next step if a solution exists for the current step. In this work, SMPC with chance constraints for an unbounded uncertainty is applied. As this setup allows constraint violations, given the risk parameter $p$, recursive feasibility cannot be proved here. The unbounded uncertainty within the coarse model was chosen to roughly over-approximate the bounded uncertainty of the detailed model. However, defining a bounded uncertainty for the coarse model could potentially yield a recursively feasible granularity R+SMPC method by applying a different SMPC method, e.g., \cite{LorenzenEtalAllgoewer2017}.

\section{Results}
\label{sec:results}
\begin{figure}
	\includegraphics[width=\columnwidth,trim=0cm 0.8cm 0cm 0cm]{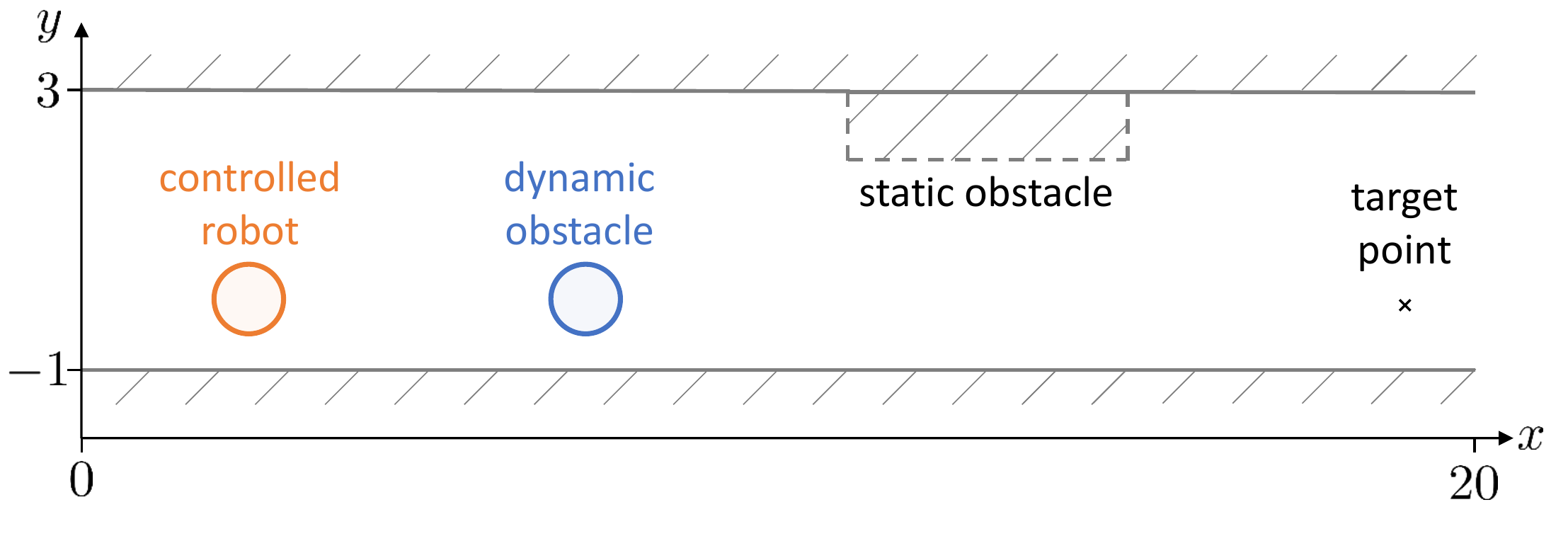}
	\caption[Simulation setup]{Simulation scenario with the controlled robot, the obstacles, and the target point for the controlled robot.}
	\label{fig:setup}
\end{figure}

In this section, we use the previously introduced method to control the motion of a mobile robot through a known landscape which consists of boundaries, a dynamic obstacle, and a static obstacle representing a narrowing road. The robot and the dynamic obstacle both have a radius of $0.5$. The objective for the controlled robot is to get from the starting point $\bm{p}_\text{start} = (0,0)$ to the target point $\bm{p}_\text{target}~=~(19,0)$ without colliding with any obstacles.
All stated values and axis types are given in SI units. The simulation scenario is shown in Figure \ref{fig:setup}.
\subsection{Simulation Setup}
The controlled robot knows the velocity and starting position of the dynamic obstacle, which starts at $\bm{p}^{\text{OB}}_{\text{start}} = (6,0)$ with constant velocity in x-direction of $v^{\text{OB}}_{x} = 0.6$. In every step, there exists an additional uncertainty for velocities in x- and y-direction, $d_{v_x}$ and $d_{v_y}$, respectively. The uncertainties are bounded by $\abs{d_{v_x}} \leq 0.1$ and $\abs{d_{v_y}} \leq 0.1$.\\
The proposed granularity MPC method requires two models. We consider a setup similar to \cite{BaethgeLuciaFindeisen2016}, which consists of two linear models for the robot. \\ 
The detailed model is given by, cf. \eqref{eq:detailed},
\begin{equation}
    \bm{x}_{k+1} = \mat{1& \Delta t & 0 & 0 \\ 0 & 1 & 0 & 0 \\ 0 & 0 & 1 & \Delta t \\ 0 & 0 & 0 & 1} \bm{x}_k+\mat{0.5{\Delta t}^2 & 0 \\ \Delta t & 0 \\ 0 & 0.5{\Delta t}^2 \\0 & \Delta t} \bm{u}_k + \bm{d}_k, \IEEEyessubnumber \label{eq:detailed_sim}\\
\end{equation}
with the sampling time $\Delta t = 0.2$ and the states $\bm{x}_k = \left(p_{x,k}, v_{x,k}, p_{y,k}, v_{y,k}\right)^\top$ consisting of the position $\left(p_{x,k},p_{y,k}\right)$ in the x-y-plane and the velocities $v_{x,k}$, $v_{y,k}$ in both directions, as well as the inputs $\bm{u}_k = \left(a_{x,k},a_{y,k}\right)^\top$ representing the acceleration in both directions. The disturbance $\bm{d}_k$ is bounded by the set $\mathbb{D} = \{\bm{d}_k \hspace*{5pt}| \hspace*{5pt}|\bm{d}_k|_{\infty} \le 0.1\}$, accounting for $d_{v_x}$ and $d_{v_y}$.
Actuator limitations are taken into account by box-constraints, bounding the inputs $\bm{u}_k$ by $|a_{x,k}| \le 3$ and $|a_{y,k}| \le 3$. The lateral position is constrained by $-0.5 \le p_{y,k} \le 2.5$ to ensure that the robot with radius $0.5$ does not leave the landscape boundaries. The velocities are constrained by $|v_{x,k}| \le 3$ and $|v_{y,k}| \le 3$.\\
A safety constraint is implemented to avoid collision with the dynamic obstacle by defining a region around the center of the dynamic obstacle in form of an ellipse \citep{BruedigamEtalWollherr2018b}. Constraint satisfaction, i.e., collision avoidance, is defined by the safety distance
\begin{equation}
    g_k = \frac{\left(p_{x,k} - p_{x,k}^{\text{OB}}\right)^2}{a^2} + \frac{\left(p_{y,k} - p_{y,k}^{\text{OB}}\right)^2}{b^2} - 1 \ge 0, \label{eq:ell_con}
\end{equation}
with the ellipse parameters $a=b=1$ and the obstacle position $\left( p_{x,k}^{\text{OB}}, p_{y,k}^{\text{OB}} \right)$. If $g_k \ge 0$ holds, the center of the robot is outside of the ellipse and the constraint is satisfied. The corners of the static box-obstacle are at the points (11,3), (11,2), (15,2), and (15,3), resulting in linear inequality constraints. \\
In order to determine the tightened constraints for the RMPC approach in the short-term prediction, we compute a disturbance invariant, outer approximation of the minimal disturbance invariant set $\mathbb{Z}$ as described in \cite{RacovicEtalMayne04} using the Multi-Parametric Toolbox~3 \citep{HercegEtalMorari2013} in MATLAB.\\
The resulting constraints for the nominal inputs $\ol{\bm{u}}_k$ are $|\ol{a}_{x,k}| \le 1.73$ and $|\ol{a}_{y,k}| \le 1.73$, while the constraints for the state $\ol{\bm{x}}_k$ of the nominal system are given by $-0.22 \le \ol{p}_{y,k} \le 2.22$, $|\ol{v}_{x,k}| \le 2.26$, and $|\ol{v}_{y,k}| \le 2.26$. 
The tighter safety constraint is challenging to compute, due to the nonlinearity of \eqref{eq:ell_con}. For simplicity, we approximated this constraint by an ellipse similar to \eqref{eq:ell_con}, enlarged by the maximal distance between the boundary of the tube and its center. The resulting ellipse parameters are $\ol{a} = \ol{b} = 2.10$. The corner points of the robust box-obstacle constraint are (10.2,3.8),(10.2,1.2),(15.8,1.2),(15.8,3.8).\\
Furthermore, the coarse model of the robot is given by, cf.~\eqref{eq:coarse},
\begin{equation}
    \bm{\xi}_{k+1} = \mat{1 & 0 \\ 0 & 1} \bm{\xi}_k+\mat{\Delta t & 0 \\0 & \Delta t} \bm{v}_k + \mat{1 & 0 \\ 0 & 1}\bm{w}_k, \IEEEyessubnumber \label{eq:coarse_sim} 
\end{equation}
where the states $\bm{\xi}_{k} = \left(p_{x,k},p_{y,k}\right)^\top$ consider only the position in the x-y-plane, and the velocities in both directions are treated as the inputs, i.e., $\bm{v}_k = \left(v_{x,k},v_{y,k}\right)^\top$. The random disturbance $\bm{w}_k$ is zero mean, normally distributed with covariance matrix $\ma{\Sigma}^w = \text{diag}(0.1,0.1)$. The risk parameter is chosen to be $p=0.8$.\\
The projection function which maps the states and inputs of the detailed model to the states and inputs of the coarse model results in
\begin{equation}
\mat{\bm{\xi}_k\\ \bm{v}_k} = \text{Proj}\left(\mat{\bm{x}_k\\\bm{u}_k}\right) = \mat{1 & 0 & 0 & 0 & 0 & 0 \\ 0 & 0 & 1 & 0 & 0 & 0 \\ 0 & 1 & 0 & 0 & 0 & 0 \\ 0 & 0 & 0 & 1 & 0 & 0} \mat{\bm{x}_k\\\bm{u}_k} .
\end{equation}
In order to maintain consistency between the models, additional input constraints are required for the coarse model that take the actuator-limitations into account. The dynamic equation for the longitudinal velocity $v_{x,k}$ in \eqref{eq:detailed_sim} is given by
\begin{equation}
    v_{x,k+1} = v_{x,k} + a_{x,k} \Delta t.
\end{equation}
Considering the limitations for the acceleration, additional input constraints for the coarse model are given by $|v_{x,k}~-~v_{x,k-1}|~\le~3 \Delta t$ and $|v_{y,k}-v_{y,k-1}| \le 3 \Delta t$, resulting from the input constraints for $\bm{u}_k$. The chance constraint to avoid the obstacles are obtained as described in Section~\ref{sec:smpc}.\\
For the \gls{OCP}, we use the quadratic running cost functions 
\begin{IEEEeqnarray}{rl}
\IEEEyesnumber
l \left(\bm{x}_k,\bm{u}_k\right) &= ||\bm{x}_k - \tilde{\bm{p}}_\text{target}||_{\bm Q} + ||\bm{u}_k||_{\bm R},\IEEEyessubnumber \\
l_{\text{c}}\left(\bm{z}_k, \bm{v}_k\right) &= ||\bm{z}_k- \bm{p}_\text{target}||_{\bm Q_{\text{c}}} + ||\bm{v}_k||_{\bm R_{\text{c}}},\IEEEyessubnumber
\end{IEEEeqnarray}
with targets $\tilde{\bm{p}}_\text{target} = (19,0,0,0)$ and $\bm{p}_\text{target} = (19,0)$, the quadratic terminal cost function $V_{\text{f,c}}\left(\bm{z}_k\right) = ||\bm{z}_k||_{\bm Q_{\text{c}}}$, and $\norm{\bm{x}}_{\bm{Q}} := \bm{x}^\top \bm{Q} \bm{x}$.
The weighting matrices are defined as $\bm{Q} = \text{diag}\left(1,0.1,1,0.1\right)$ and $\bm{R}=\text{diag}\left(0.1,0.1\right)$, as well as $\bm{Q}_{\text{c}}=\text{diag}\left(1,1\right)$ and $\bm{R}_{\text{c}} = \text{diag}\left(0.1,0.1\right)$. The feedback gains for the detailed and the coarse model are
\begin{equation}
K = \mat{3.77 & 4.67 & 0 & 0 \\ 0 & 0 & 3.77 & 4.67} , K_{\text{c}} = \mat{2.32 & 0 \\ 0 & 4.14}.
\end{equation}
We choose the short-term horizon $N_{\text{s}}=7$ and the long-term horizon $N_{\text{l}}=13$, resulting in $N = 20$.

\subsection{Simulation Results}
In the following we will first analyze the behavior of the robot and then evaluate the performance, i.e., cost, and computational effort of the proposed method. To get an appropriate comparison, we implement three different methods to evaluate and compare our proposed control scheme.
\begin{enumerate}
	\item Granularity R+SMPC: The method proposed in this work uses RMPC with the detailed model for short-term predictions and SMPC with the coarse model for long-term predictions.
	\item Single model R+SMPC: This method uses RMPC with the detailed model for  short-term predictions and SMPC with the same detailed model for long-term predictions.
	\item Single model RMPC: This method uses RMPC with the detailed model for the entire prediction horizon.
\end{enumerate}
Each method is simulated 100 times. The simulations were carried out in MATLAB using the \textit{fmincon} solver on a standard desktop computer.

\subsubsection{Controlled robot behavior.}
\begin{figure}
	\includegraphics[width=\columnwidth,trim=0.5cm 4.7cm 0.5cm 1.5cm]{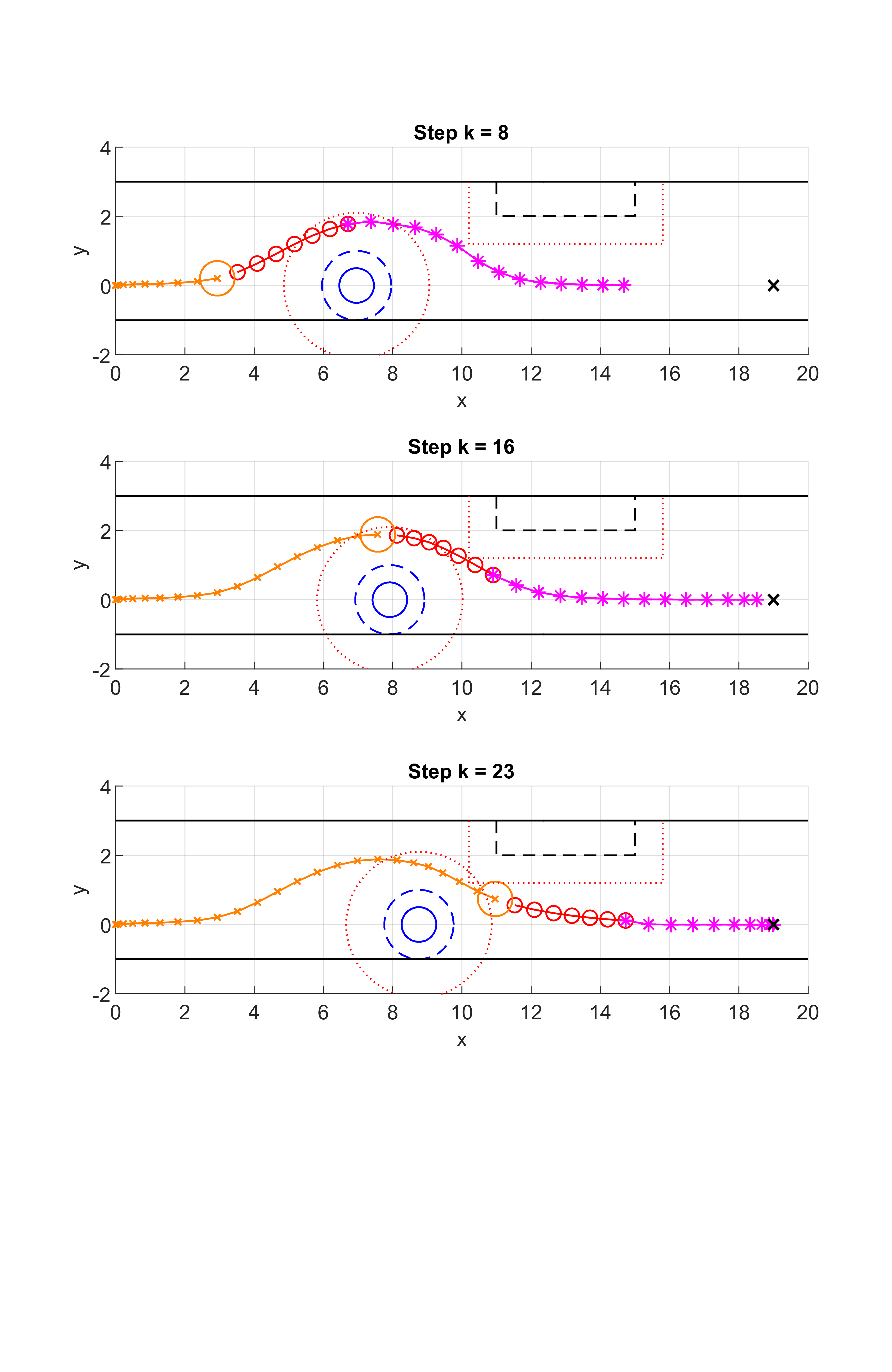}
	\caption[Simulation setup]{Simulation results of the proposed granularity R+SMPC method. Previous robot steps are shown in orange. The planned RMPC trajectory is shown in red circles, the planned SMPC trajectory in purple asterisks. The constraints for the nominal states of the RMPC are displayed by dotted red lines. Note that the nominal trajectory is omitted for clarity, only the planned RMPC trajectory is displayed. }
	\label{fig:granRS}
\end{figure}

\begin{figure}
	\includegraphics[width=\columnwidth,trim=1cm 1.9cm 1cm 1.5cm]{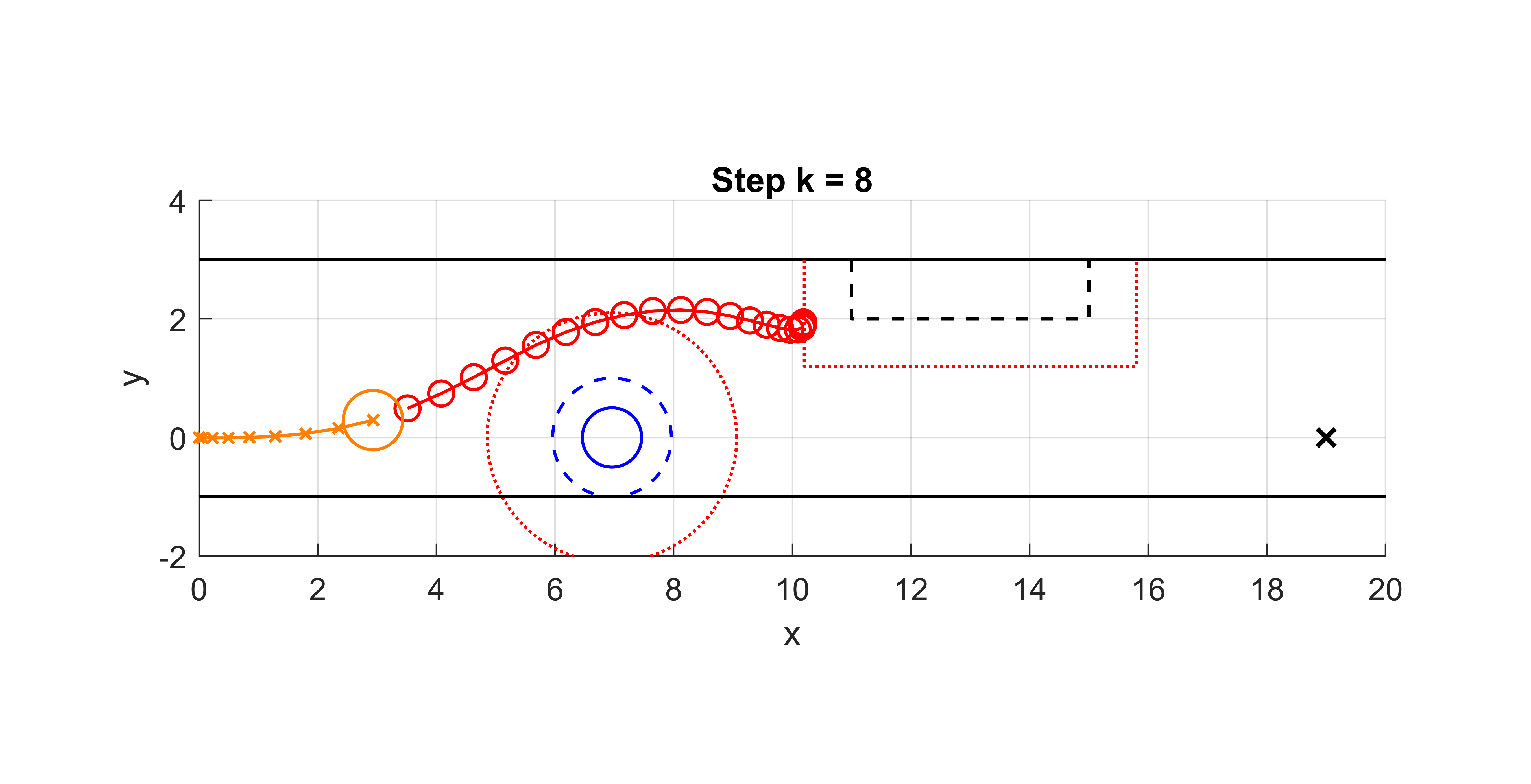}
	\caption[Simulation setup]{Simulation results of RMPC on a detailed model for the entire horizon. Previous robot steps are shown in orange. The planned RMPC trajectory is shown in red circles. The constraints for the RMPC are displayed by dotted red lines.}
	\label{fig:oneR}
\end{figure}

The results for one example simulation of the proposed granularity R+SMPC method are displayed in Figure~\ref{fig:granRS}.
Collisions are avoided if the center of the controlled robot lies outside the dashed blue circle. At step $k=8$ the robot is planning to pass the moving obstacle, while avoiding the static obstacle. The dotted red lines are the obstacle constraints considered for RMPC. It is to note that the constraint for the dynamic obstacle moves in each prediction step, while in Figure~\ref{fig:granRS} only the constraint for the current dynamic obstacle position is displayed. The constraints for further predicted dynamic obstacle positions are omitted for better visibility. 
Therefore, the planned RMPC trajectory only avoids the displayed dynamic obstacle constraints in the current step shown. It is sufficient that the nominal trajectory satisfies the robust constraint, cf. Section \ref{sec:impRMPC}. 
In other words, the current state still satisfies the robust constraint even if it is inside the dotted red circle, as long as its corresponding nominal state satisfies the robust constraint. In Figure \ref{fig:granRS} the nominal trajectory is omitted for clarity.\\
The planned SMPC trajectory does not consider the RMPC constraints for obstacles and boundaries, but satisfies chance constraints, which allows the robot to plan passing the dynamic obstacle. Steps $k=16$ and $k=23$ show that the robot successfully moves around the dynamic obstacle. In all 100 simulations the controlled robot successfully passes the dynamic obstacle without feasibility issues of the \gls{OCP}.\\
We now compare the behavior of the proposed method to the two other approaches. First, using the single model R+SMPC approach yields a similar trajectory to the previously shown simulation. Therefore, a display of the results is omitted. \\
Second, we evaluate the results of applying the single model RMPC approach. The controlled robot does not pass the dynamic obstacle in any of the 100 simulations. Step $k=8$ of a sample simulation is shown in Figure~\ref{fig:oneR}. Unlike the first simulation with the proposed method, the robust approach is more conservative and fails to pass the dynamic obstacle. The robust constraints are enforced on the entire horizon, forcing the robot to stop in front of the static box-obstacle in order to avoid potential constraint violations. 

\subsubsection{Cost and computational effort.}

\begin{figure}
	\includegraphics[width=\columnwidth,trim=0cm 0.5cm 0cm 0cm]{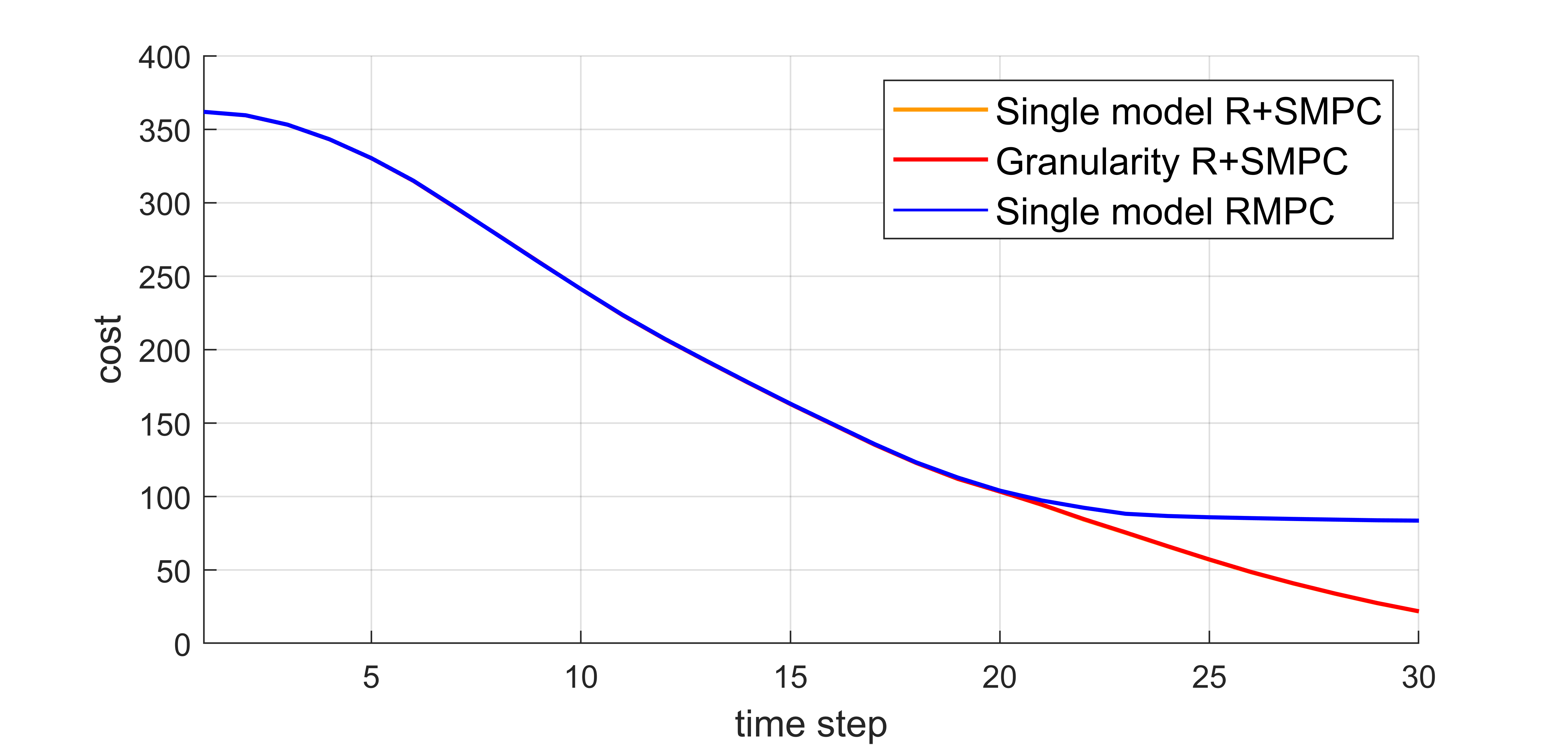}
	\caption[Cost]{Cost comparison of the three approaches. Granularity R+SMPC and single model R+SMPC yield similar cost.}
	\label{fig:cost}
\end{figure}

\begin{figure}
	\includegraphics[width=\columnwidth,trim=0cm 0.5cm 0cm 0.6cm]{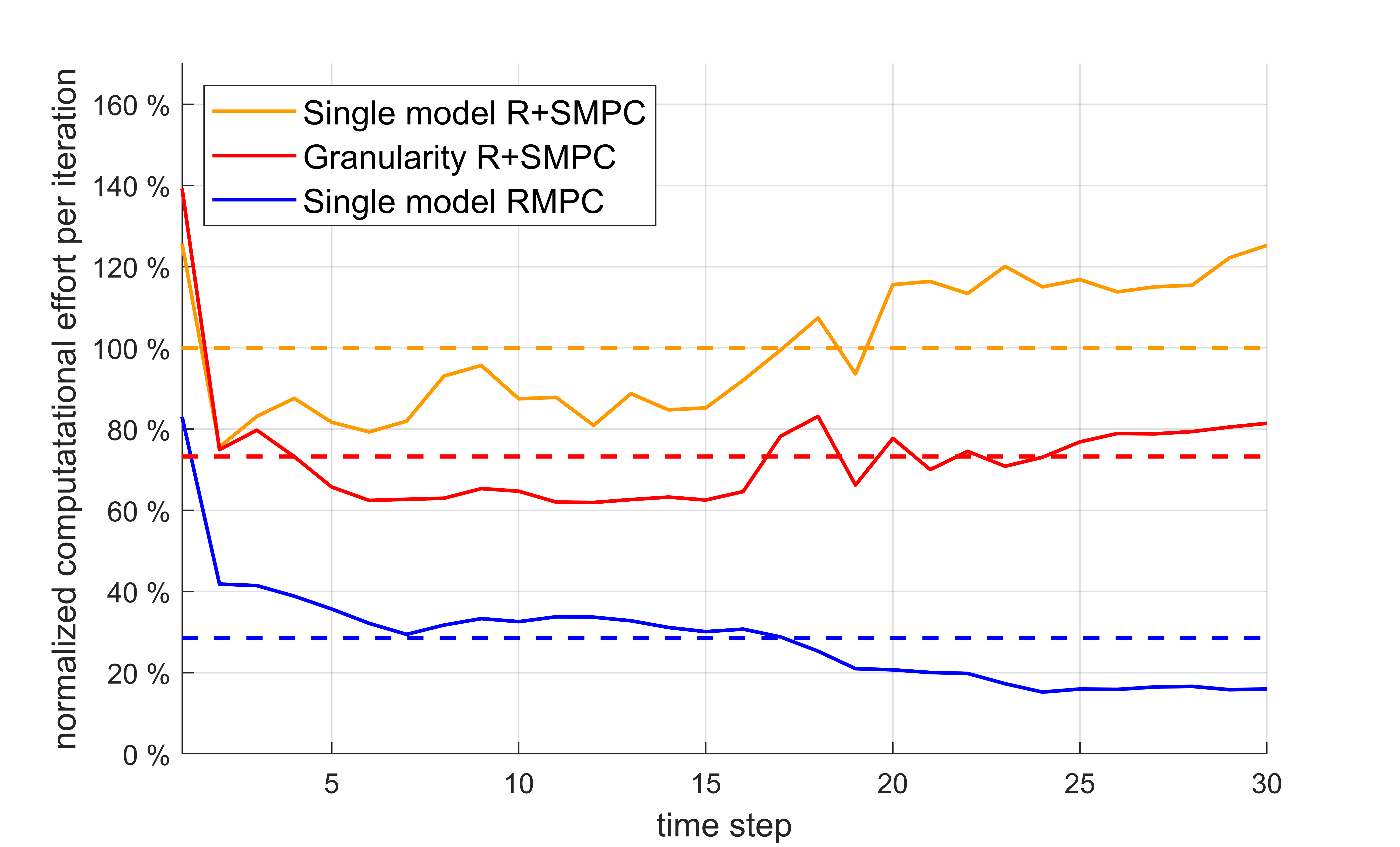}
	\caption[Computation]{Computational effort of the three approaches. Dashed lines represent mean values over all steps. The mean value of the computational effort for the single model R+SMPC is chosen as the base value.}
	\label{fig:timeperiter}
\end{figure}

After having analyzed the behavior of the proposed method, the cost and computational effort is now evaluated. For each time step a mean value over the 100 simulations is calculated for cost and computational effort.\\
We compare the costs of the three approaches, by evaluating, at each step, the cost function
\begin{equation}
    l \left(\bm{x}_k,\bm{u}_k\right) = ||\bm{x}_k-\tilde{\bm{p}}_\text{target}||_{\bm Q} + ||\bm{u}_k||_{\bm R}.
\end{equation}
Figure \ref{fig:cost} shows that the cost for the single model R+SMPC is equal to the cost of the proposed granularity R+SMPC method. 
Applying the single model RMPC method results in a more conservative robot behavior, i.e., the robot does not pass the dynamic obstacle in front of the static box-obstacle. This can also be seen in the costs, which are similar to the granularity R+SMPC method at first, but then remain at a higher level as the robot is unable to move closer to the target point.\\
We now evaluate the computation time to solve the optimization problems at each time step. For the comparison of computational effort, the mean value over all steps of the single model R+SMPC approach is chosen as the base value, with a computation time of $\SI{3.2}{\second}$. The results are displayed in Figure \ref{fig:timeperiter}.
For this simulation setup, the computational effort of the proposed granularity R+SMPC method is $73\%$ of the single model R+SMPC approach. Therefore, by using a simple model for the remote horizon, the computational effort can, on average, be reduced by~$27\%$. The computational effort of the single model RMPC approach is even lower, as no chance constraints are considered. However, as shown before, the solution is more conservative.\\
In summary, for this specific simulation the proposed granularity R+SMPC method results in less computational effort compared to a single model R+SMPC approach, while the performance, evaluated by the cost function, remains similar. The proposed approach is less conservative compared to a single model RMPC approach. 


\section{Conclusion}
\label{sec:conclusion}

In this work, we proposed a mixed RMPC and SMPC method which uses two models for the prediction horizon, a detailed model for short-term predictions and a coarse model for long-term predictions. RMPC is used with a detailed model, while chance constrained SMPC is combined with a coarse model. In a simulation study the proposed approach yields lower computational effort compared to a combined RMPC and SMPC method with a single detailed prediction model, while yielding similar cost.\\
The proposed approach allows to robustly plan short-term trajectories, while considering long-term targets with the SMPC approach using the coarse model. This is advantageous as precise long-term predictions are often challenging, resulting in overly conservative RMPC trajectories and unnecessary model complexity. While standard RMPC approaches require a trade-off between model accuracy, horizon length, and computation time, the proposed approach enables easier adaptation of the MPC problem to specific tasks. This can be beneficial in various application, e.g., autonomous driving.


\bibliography{Dissertation_bib}
                                                   







\appendix

\end{document}